\documentclass[%
 reprint,
superscriptaddress,
 amsmath,amssymb,
 aps,
]{revtex4-2}

\usepackage{graphicx}
\usepackage{dcolumn}
\usepackage{bm}
\usepackage{color}

\begin{document}


\title{Constructing the Fulde–Ferrell–Larkin–Ovchinnikov state in antiferromagnetic insulator CrOCl}

\author{Yifan Ding}
\homepage{These authors contributed equally to this work.}
\author{Jiadian He}
\homepage{These authors contributed equally to this work.}
\affiliation{School of Physical Science and Technology, ShanghaiTech University, Shanghai 201210, China}
\affiliation{ShanghaiTech Laboratory for Topological Physics, ShanghaiTech University, Shanghai 201210, China}

\author{Shihao Zhang}
\homepage{These authors contributed equally to this work.}
\affiliation{School of Physical Science and Technology, ShanghaiTech University, Shanghai 201210, China}
\affiliation{School of Physics and Electronics, Hunan University, Changsha 410082, China}

\author{Huakun Zuo}
\homepage{These authors contributed equally to this work.}
\affiliation{Wuhan National High Magnetic Field Center, Huazhong University of Science and Technology, Wuhan 430074, China}

\author{Pingfan Gu}
\affiliation{State Key Laboratory for Mesoscopic Physics, Nanooptoelectronics Frontier Center of the Ministry of Education,  School of Physics, Peking University, Beijing 100871, China}

\author{Jiliang Cai}
\author{Xiaohui Zeng}
\affiliation{School of Physical Science and Technology, ShanghaiTech University, Shanghai 201210, China}
\affiliation{ShanghaiTech Laboratory for Topological Physics, ShanghaiTech University, Shanghai 201210, China}

\author{Pu Yan}
\author{Kecheng Cao}
\affiliation{School of Physical Science and Technology, ShanghaiTech University, Shanghai 201210, China}

\author{Kenji Watanabe}
\affiliation{Research Center for Functional Materials, National Institute for Materials Science, Tsukuba 305-0044, Japan}
\author{Takashi Taniguchi}
\affiliation{International Center for Materials Nanoarchitectonics, National Institute for Materials Science, Tsukuba 305-0044, Japan}

\author{Peng Dong}
\author{Yiwen Zhang}
\author{Yueshen Wu}
\author{Xiang Zhou}
\author{Jinghui Wang}
\email{wangjh2@shanghaitech.edu.cn}
\affiliation{School of Physical Science and Technology, ShanghaiTech University, Shanghai 201210, China}
\affiliation{ShanghaiTech Laboratory for Topological Physics, ShanghaiTech University, Shanghai 201210, China}

\author{Yulin Chen}
\affiliation{School of Physical Science and Technology, ShanghaiTech University, Shanghai 201210, China}
\affiliation{ShanghaiTech Laboratory for Topological Physics, ShanghaiTech University, Shanghai 201210, China}
\affiliation{Department of Physics, Clarendon Laboratory, University of Oxford, Oxford OX1 3PU, UK}

\author{Yu Ye}
\email{ye\_yu@pku.edu.cn}
\affiliation{State Key Laboratory for Mesoscopic Physics, Nanooptoelectronics Frontier Center of the Ministry of Education,  School of Physics, Peking University, Beijing 100871, China}

\author{Jianpeng Liu}
\email{liujp@shanghaitech.edu.cn}
\author{Jun Li}
\email{lijun3@shanghaitech.edu.cn}
\affiliation{School of Physical Science and Technology, ShanghaiTech University, Shanghai 201210, China}
\affiliation{ShanghaiTech Laboratory for Topological Physics, ShanghaiTech University, Shanghai 201210, China}




\date{\today}

\begin{abstract}
Time reversal symmetry breaking in superconductors, resulting from external magnetic fields or spontaneous magnetization, often leads to unconventional superconducting properties. In this way, a conventional Fulde-Ferrell-Larkin-Ovchinnikov (FFLO) state, characterized by the Cooper pairs with nonzero total momentum, may be realized by the Zeeman effect caused from external magnetic fields. Here, we report the observation of superconductivity in a few-layer antiferromagnetic insulator CrOCl by utilizing superconducting proximity effect with NbSe$_2$ flakes. The superconductivity demonstrates a considerably weak gap of about 0.12 meV and the in-plane upper critical field reveals as behavior of the FFLO state at low temperature. Our first-principles calculations indicate that the proximitized superconductivity may exist in the CrOCl layer with Cr vacancies or line-defects. Moreover, the FFLO state could be induced by the inherent larger spin splitting in the CrOCl layer. Our findings not only demonstrate the fascinating interaction between superconductivity and magnetism, but also provide a possible path to construct FFLO state by intrinsic time reversal symmetry breaking and superconducting proximity effect.

\end{abstract}

\maketitle


\section{INTRODUCTION}


Symmetry breaking, one of the core problems in condensed matter physics in recent decades, plays an important role in the development of unconventional superconductivity \cite{TSC, Iron}. For a conventional superconductor, spatial inversion and time reversal symmetries are preserved, and the wave function of Cooper pairs follows the traditional Bardeen-Cooper-Schrieffer (BCS) mechanism (Fig. 1a) \cite{BCS}. On the one hand, for the case of time reversal symmetry breaking, the external magnetic field causes the Zeeman effect (Fig. 1b). In this case, the conventional FFLO state can be realized in a clean limit superconductor \cite{FF, LO, FFLOclean1, FFLOclean2, FFLO_PRX}, which is often observed in the layered organic materials \cite{FFLOin2D, Organic}, heavy-fermion superconductors \cite{Heavy3, Heavy1, Heavy2}, iron-based superconductors \cite{KFeAs, FeSe,FeSe2}, transition metal dichalcogenides \cite{NbS2} and Sr$_2$RuO$_4$ \cite{SRO}. There, the order parameter of the finite-momentum Cooper pairs can be periodic modulated in real space at low temperatures. On the other hand, for the case of spatial inversion symmetry breaking, the presence of Rashba-type or Ising-type spin-orbital coupling (SOC) in a superconductor can enhances the in-plane upper critical field beyond Pauli limit $B_p$=1.86 $T_c$ (Fig. 1c) \cite{non-centrosymmetric, Ising, Crossover}. Besides, for the case of simultaneously breaking the spatial inversion and time reversal symmetries, Rashba-type or Ising-type FFLO state has also been studied (Fig. 1d) \cite{FFLOPNAS, Orbital, FFLOinIsing, FFLOinRashba}. Therefore, it is commonly believed that time reversal symmetry breaking represents a crucial prerequisite for achieving unconventional superconductivity and a robust spin structure is necessary for the FFLO state. 


\begin{figure*}[b]
\centering
\includegraphics[width=1.0\linewidth]{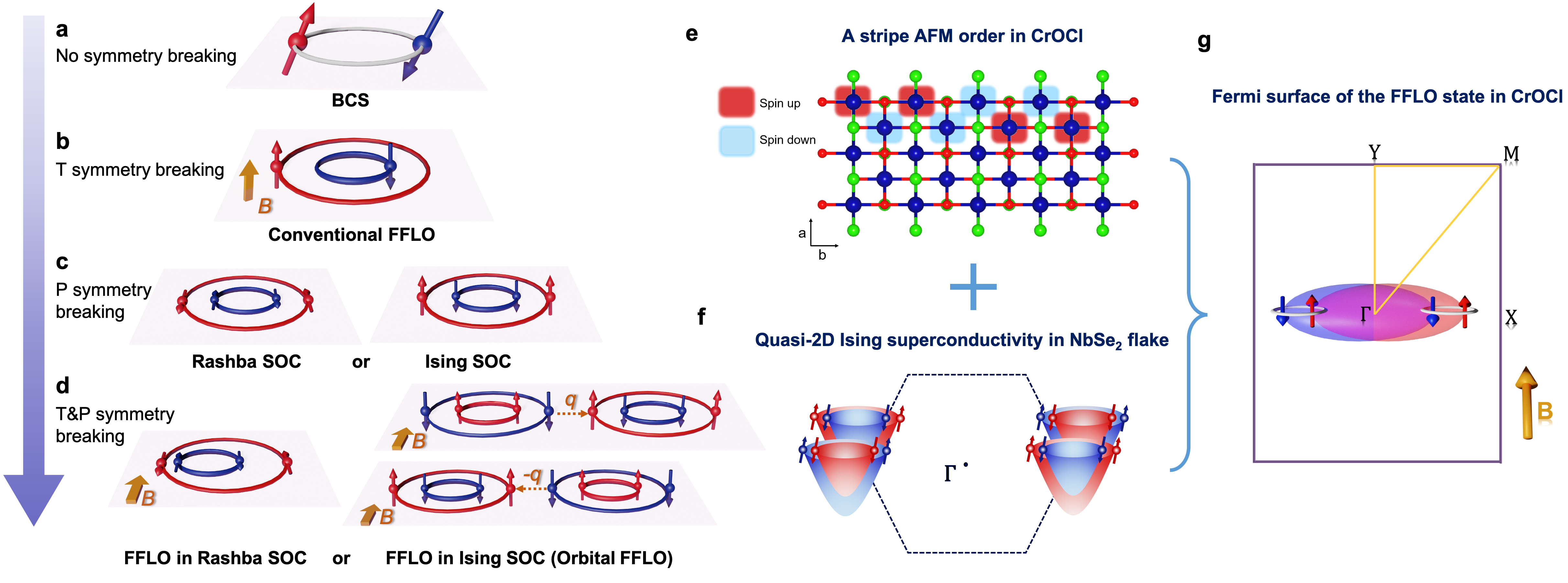}
\caption{{\label{fig1} \textbf{Symmetry breaking and pairing states in superconductors.} \textbf{(a)} A traditional BCS pairing without symmetry breaking. \textbf{(b)} A conventional FFLO state with time reversal symmetry breaking. \textbf{(c)} Rashba-type SOC or Ising-type SOC with spatial inversion symmetry breaking. \textbf{(d)} Unconventional FFLO  state in Rashba-type SOC or Ising-type SOC with both breaking the spatial inversion and time reversal symmetries. \textbf{(e)} The stripy AFM order of CrOCl, leads to the intrinsic time reversal symmetry breaking. \textbf{(f)} Schematic of the pairing symmetry in a NbSe$_2$ flake. The properties of the NbSe$_2$ flake are close to that of bulk, which can be considered that the spatial inversion symmetry is preserved. \textbf{(g)} The simplified illustration of FFLO state formation in few-layer CrOCl, where the in-plane magnetic field splits the spin degeneracy at $\Gamma$ point.}}
\label{fig:figs1}
\end{figure*}

Recently, a van der Waals (vdW) insulator CrOCl has attracted significant attention due to its novel antiferromagnetic (AFM) order \cite{CrOCl1975, CrOCl2019,CrOCl,CrOCl2022,CrOCl2023}. The antiferromagnetism transition of CrOCl occurs at $T_N$=13.5 K as confirmed by magnetic susceptibility measurements, and such first-order transition is also accompanied by an orthorhombic to monoclinic lattice distortion \cite{CrOCl2009}. In addition, the ground state of CrOCl is found as an unconventional spin-density wave (SDW) state along the short axis, namely, a stripy AFM order (Fig. 1e) \cite{CrOCl}. Although CrOCl behaves as an insulator with a giant band gap \cite{PRM}, the structural distortion, magnetic and electronic ordering resemble the parent compound of iron-based superconductors. Therefore, it is highly possible to observe exotic superconducting state by inducing carrier doping into CrOCl. Despite significant efforts in chemical doping, CrOCl remains a challenging material due to the large electronegativity of Cl$^-$ ions, which poses a considerable obstacle in terms of carrier trapping or exclusion \cite{doping}. Moreover, modifying the band structure of chlorides to achieve a conducting state through high pressure or electric field gating seems to be a difficult task \cite{pressure}. Thus, we focus on the realization of superconductivity in CrOCl by the superconducting proximity effect.

Superconducting proximity effect has been widely studied in recent decades and shown great significance in the area of topological superconductivity \cite{Kane}. When a superconducting material and a non-superconducting material come into contact with each other, the superconducting wave function can be induced from an even-parity $s$-wave superconductor to a non-superconducting state within a characterization coherence length $\xi_N$. Particularly, when the non-superconducting side behaves as magnetism, unconventional superconductivity or even odd-parity superconductivity may occur due to the spontaneous time reversal symmetry breaking properties \cite{Spintriplet2006, Spintriplet2010, Spintriplet2016, Spintriplet2021}. Such interplay between the time reversal symmetry breaking and the spin-singlet superconductivity may induce novel quantum phenomena, including the spin-triplet superconductivity, the time reversal invariant topological superconductors, and Majorana fermions \cite{Spintriplet2006, Spintriplet2010,  Spintriplet2016, Spintriplet2021, Topological1, Topological2}.

It is worth noting that by measuring the tunneling current through a vdW heterostructure of graphite/CrOCl/graphite, the magnetic transitions of few-layer CrOCl are similar to that of bulk \cite{CrOCl}. Since the electron wave function can be correlated into CrOCl for more than 12 layers \cite{CrOCl}, it may be a promising way to penetrate Cooper pairs into CrOCl by superconducting proximity effect as well. In the magnetism/superconductivity heterostructure, the large spin splitting is inherent for magnetic system and the carrier doping is coming from the superconductor. For the case of CrOCl, the in-plane magnetic field will break the Kramers degeneracy between the energy bands of opposite spins in CrOCl layer, and induce remarkable spin splitting. In this way, the time reversal symmetry is destroyed and conventional FFLO state may emerge. The quasi-classical picture of FFLO state reveals that the large spin splitting and slight carrier doping is necessary for the emergence of FFLO state \cite{FFLO_PRX}, where the electrons from two spin channel may form the Cooper pair with nonzero total momentum as shown in Fig.1(g).

In this work, we studied the proximity-effect-induced superconductivity with FFLO state in few-layer AFM insulator CrOCl by preparing the CrOCl/NbSe$_2$ vdW heterostructure. The device was in a specialized design to characterize transport properties of the insulator CrOCl directly. Quantum transport measurements revealed its two-dimensional (2D) superconducting properties of few-layer CrOCl. Moreover, the FFLO state was observed when the magnetic field was applied along in-plane direction. The differential resistance spectra clearly showed the proximity-effect-induced superconducting gap in the CrOCl layers. Combing the experimental results and first-principles calculations, we uncovered the origin of FFLO state in the few-layer AFM insulator CrOCl. Our work provides a possible scheme for constructing FFLO state and paves the way for exploring the interaction between antiferromagnetism and superconductivity.

\section{SUPERCONDUCTING CHARACTERIZATIONS OF DEVICE}

\begin{figure*}[b]
\centering
\includegraphics[width=0.8\linewidth]{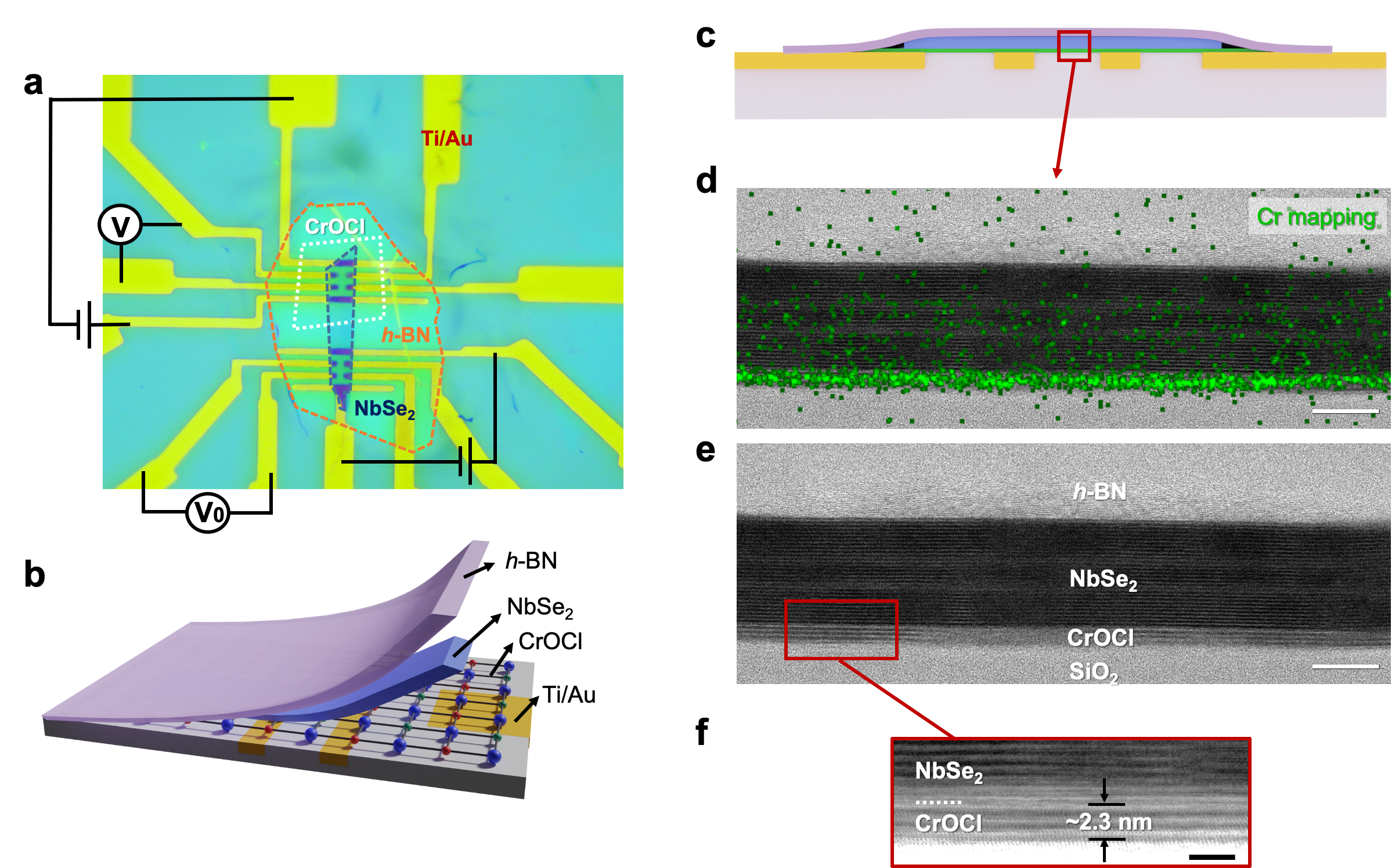}
\caption{{\label{fig2} \textbf{Structure of CrOCl/NbSe$_2$ device.} \textbf{(a)} Optical image of the device. The electrodes can be connected in different configurations for current flowing and voltage measurements to compare the electrical properties of intrinsic NbSe$_2$ ($V_0$) and CrOCl/NbSe$_2$ ($V$). The CrOCl, NbSe$_2$, and $h$-BN flakes are labeled in white, blue, and orange dot lines, respectively. \textbf{(b)} Schematic image of the stacking order of $h$-BN, NbSe$_2$, and CrOCl flakes. Thus, the electrodes are attached to the CrOCl layer instead of NbSe$_2$. \textbf{(c)} Cross-sectional view of the CrOCl/NbSe$_2$ heterostructure.  \textbf{(d)} Transmission electron microscopy (TEM) image of the cross-section of the heterostructure, and the corresponding elemental mapping of Cr by energy dispersive x-ray analysis. A few atomic layers of Cr distribution can be observed as the green dot concentration region. The scale bar is 20 nm. \textbf{(e)} The high-resolution TEM image of the heterostructure demonstrates the layered structure of CrOCl and NbSe$_2$ crystals, and \textbf{(f)}  the enlarged view of the interface region. The white dash line reveals the interface between CrOCl and NbSe$_2$. The thickness of CrOCl is about 2.3 nm, which corresponds to 3 atomic layers. }}
\label{fig:fig2}
\end{figure*}

\begin{figure*}[b]
\centering
\includegraphics[width=0.9\linewidth]{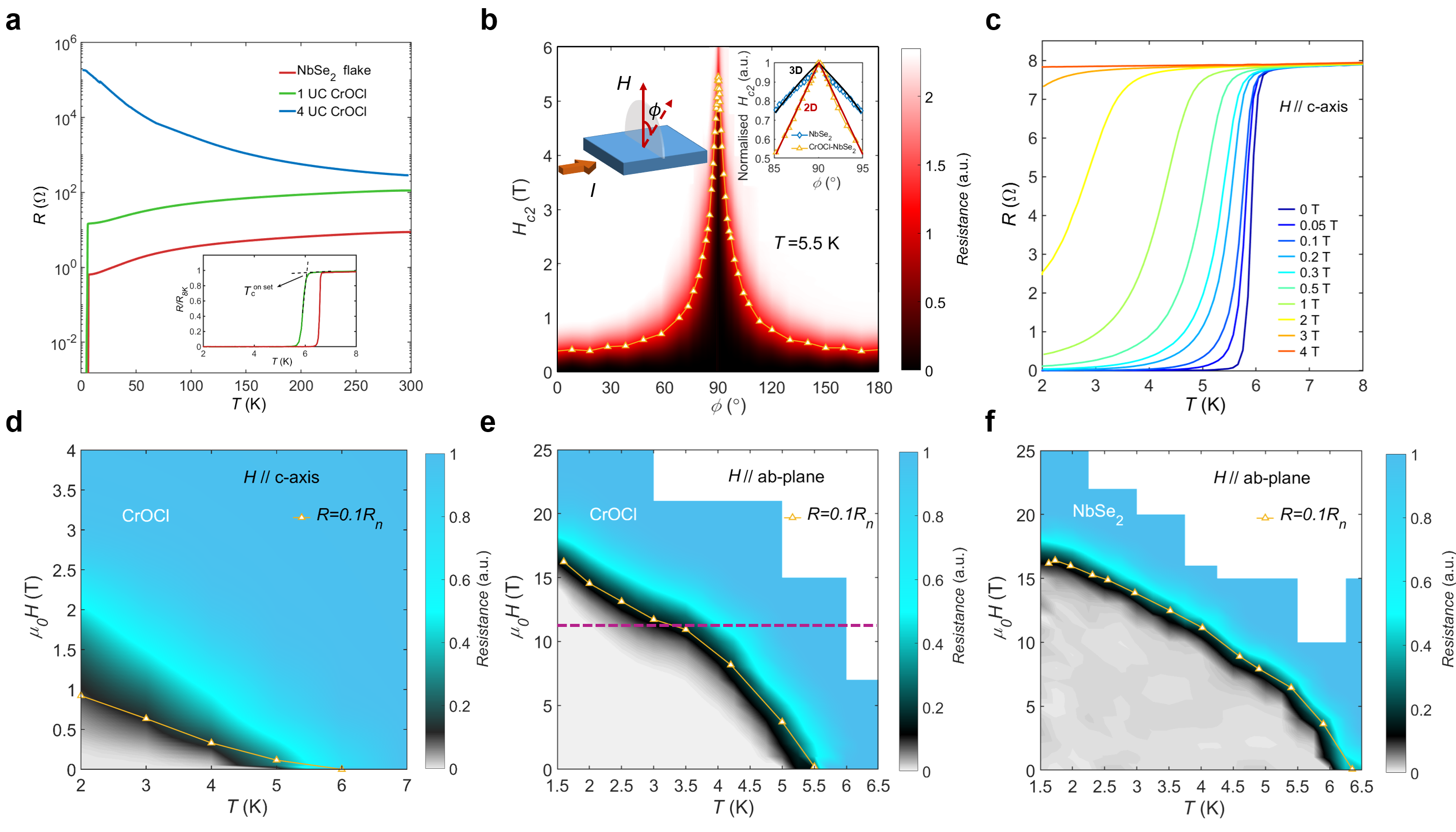}
\caption{{\label{fig3} \textbf{Superconducting characterizations of CrOCl/NbSe$_2$.} \textbf{(a)} Temperature dependence of resistance in CrOCl/NbSe$_2$, where CrOCl varies in different thickness at 1 UCs and 4 UCs. The inset shows the resistance ($R/R_{8K}$) in the range from 2 K to 8 K, where $T_{c}^{on set}$ marks the onset of superconducting transition of CrOCl. \textbf{(b)} The angular dependence of $H_{c2}$ with fields rotating along out-of-plane. Here $\phi$ demonstrates the angle between $H$ and $ab$-plane. The $H_{c2}$ is defined as the half of normal resistance. The $H_{c2}$-$\phi$  is fitted by both 2D Tinkham model (red) and 3D anisotropic Ginzburg-Landau model (blue), respectively. \textbf{(c)} Resistance of monolayer-CrOCl/NbSe$_2$ as a function of temperature for the out-of-plane magnetic fields, varying from 0 T to 4 T.  \textbf{(d)} Color mapping of the out-of-plane magnetoresistance for the monolayer-CrOCl/NbSe$_2$. Here, the $H_{c2}$ is considerably larger than that of $H_{c2}^{zero}$, namely, the $H_{c2}$ at completely zero resistance corresponding to the proximity-induced superconductivity of monolayer CrOCl. \textbf{(e)} The in-plane magnetoresistance color mapping for the monolayer-CrOCl/NbSe$_2$. The $H_{c2}$ data points in black solid line are fitted obeying a superconductor in the FFLO state. \textbf{(f)} Color mapping of in-plane magnetoresistance for 22-UC NbSe$_2$. Black solid line demonstrates the Ginzburg-Landau fitting for the $H_{c2}$. }}
\label{fig:fig3}
\end{figure*}

The sample was designed as a vdW heterostructure device of BN/NbSe$_2$/CrOCl as shown in the optical image in Fig. \ref{fig2}(a), and the layer directly in contact with the electrode was CrOCl. First of all, Ti/Au electrodes were embedded into the SiO$_2$/Si substrates as illustrated in Fig. \ref{fig2}(b) and the cross-sectional view is shown in Fig. \ref{fig2}(c). In this way, the CrOCl and NbSe$_2$ layers can be stacked onto a flat substrate to avoid the cracking problem which often exists at the edge steps of the electrodes in a conventional method. Otherwise, the problem of shortage cannot be circumvented easily in monolayer CrOCl, as introduced in the Supplementary Information. More importantly, the monolayer CrOCl can be well restricted from atmosphere linking through the edge steps. In this vdW heterostructure device, the interfaces between Ti/Au electrodes, CrOCl, NbSe$_2$, and $h$-BN play a key role on the electrical transport properties. In Fig. \ref{fig2}(d) and \ref{fig2}(e), the scanning transmission electron microscopy (STEM) and the energy dispersive x-ray reveal the high quality interfaces of Au-CrOCl, CrOCl-NbSe$_2$, and also NbSe$_2$-BN, for which the interfaces are in good vdW contact, and any bubble or degradation have been eliminated. Particularly, when we enlarge the interface region for CrOCl/NbSe$_2$ (see Fig. \ref{fig2}(f)), the atomic layers of both CrOCl and NbSe$_2$ can be well identified, and the as-studied CrOCl consists of three atomic layers. Thus, within the specifically structured device, the abstracted circuit can be demonstrated as shown in Fig. S7, where the current flows through both CrOCl and NbSe$_2$. Since the interface between the insulating CrOCl and the metallic NbSe$_2$ behaves as a Schottky barrier, the contact resistance $R_c$ should be considerably large. Furthermore, as mentioned above, although CrOCl exhibits large resistance as an insulator especially at low temperature, few layers of CrOCl can exhibit good electrical conductivity properties via charge or Cooper pair transfer \cite{CrOCl}.

Fig. \ref{fig3}(a) gives the temperature dependence of normalized resistance for CrOCl with different thicknesses of one and four unit cells (UCs), and the superconductivity of intrinsic NbSe$_2$ is also measured as illustrated in Fig. \ref{fig3}(a). The critical temperature ($T_c$) of the monolayer-CrOCl/NbSe$_2$ is comparable with that of the intrinsic NbSe$_2$. While as the thickness of CrOCl is up to 4 UCs, the $R$-$T$ curve demonstrates an obvious insulating behavior. The thickness of NbSe$_2$ flake in our experiments is about 15 nm ($\sim$ 22 UCs) as confirmed from the TEM measurements or atomic force microscopy, and the temperature dependence of NbSe$_2$'s resistance behaves as a bulk crystal as well studied in previous work \cite{NbSe2}. Therefore, the superconducting transition in the CrOCl layer should be correlated to the NbSe$_2$ flake based on the charge transfer and superconducting proximity effects.

Basically, the superconducting wave function from NbSe$_2$ reveals a $s$-wave pairing symmetry, in which the paired electrons carry the opposite spins, resulting in a so-called singlet Cooper pairs. When the superconductor is coupled with a normal metal, a supercurrent can flow into the metal within a distance less than a coherence length $\xi_N$ = $\sqrt{\hbar D/k_BT} $,  where $D$ is the diffusion coefficient, $\hbar$ is the Planck's constant, and $k_B$ is the Boltzmann's constant \cite{FM/SC}. Normally, the $\xi_N$ can be up to even tens of nanometers. For a magnetic material, however, the superconducting coherence length in magnetic state ($\xi_M$) turns to be $\xi_M$ = $\sqrt{\hbar D/k_BT_M} $, where $T_{M}$ is the magnetic transition temperature. As a result, the $\xi_M$ is dramatically small as about 1 nm even less for a strongly coupling ferromagnetic state, and the superconducting wave function decays exponentially $\Delta \sim$ exp($D_M/\xi_M$) \cite{FM/SC}. The major reason is that the spin polarization in the magnetic materials will easily destroy singlet Cooper pairs. Nevertheless, the superconducting coherence length of AFM state is between that of normal metals and ferromagnetic materials. For the AFM CrOCl in the present case, the diffusive penetration depth seems likely to be about 2.7 nm (3 UCs). In addition, the superconducting current flowing through the CrOCl between two NbSe$_2$ layers can be present at 6 UCs, that is, the critical thickness of Josephson coupling is about 6 UCs as provided in the Supplementary Information Fig. S11. Therefore, we can basically conclude that the diffusive penetration depth of CrOCl is about 3 UCs as $\sim$2.7 nm, in which such proximity-effect-induced superconductivity should indicate low-dimensional nature.

To investigate the possible two-dimensional superconductivity, we then studied the angular dependence of upper critical fields ($H_{c2}$) in monolayer-CrOCl/NbSe$_2$ as shown in Fig.\ref{fig3}(b). The $H_{c2}$ along the $ab$-plane ($H_{c2}^{ab}$ $\sim$6 T at 5.5 K) is dramatically larger than that of $c$-axis $H_{c2}^{c}$ ($\sim$0.2 T at 5.5 K). And particularly, the angular dependence of $H_{c2}$ around $ab$-plane ($\phi$=90$^{\circ}$) reveals as a 2D Tinkham model rather than the 3D anisotropic Ginzburg-Landau (GL) model, indicating that the proximity-induced superconductivity behaves as a 2D superconductor, being well consistent with field-effected MoS$_2$ \cite{MoS2_1, MoS2_2}, $1T_{d}$-MoTe$_2$ \cite{MoTe2}, and heterointerface of NbSe$_2$ and CrCl$_3$ \cite{NSCrCl}. In contrast, the angular dependence of $H_{c2}$ in NbSe$_2$ obeys the 3D anisotropic Ginzburg-Landau model.

We also studied the magnetoresistance of monolayer-CrOCl/NbSe$_2$ along out-of-plane and in-plane field. When the magnetic field is applied along out-of-plane direction (see Fig. \ref{fig3}(c) and \ref{fig3}(d)), the proximity-induced superconductivity can be suppressed by a considerably weak magnetic field as $\sim$~ 0.2 T at 2 K which corresponds to the $H_{c2}^{zero}$, and it is far more less than that of intrinsic NbSe$_2$ ($\sim$~ 2 T at 2 K). However, when an in-plane magnetic field is applied (see Fig. \ref{fig3}(e)), both proximity-induced superconductivity of CrOCl and intrinsic superconductivity of NbSe$_2$ indicate a strong diamagnetism, for which we applied pulsed high magnetic fields up to 30 T to evaluate the $H_{c2}$ as given in Fig. S9. Particularly, the $H_{c2}^{ab}$ of monolayer CrOCl is large as $\sim$ 18.23 T at 1.6 K, which is even beyond the Pauli paramagnetic limit ($H_{c2}^{p}$ = 13.7 T), and the $H_{c2}^{ab}$ shows an obvious upturn profile at the low temperature regions, resulting in a kink at 3.5 K. Such temperature dependence behavior of $H_{c2}^{ab}$ is a hallmark of FFLO state, which is different from the Ginzburg-Landau formula for the conventional superconductors \cite{Ising, NbS2, Abrikosov}. In this FFLO state, the observed coherence length ($\xi$) can be estimated to be about 2 nm from the $H_{c2}$ results based on the 2D Tinkham's model. Thus, this small coherence length contributes the formation of FFLO state \cite{FFLO_PRX}.


Although strong SOC in few-layer NbSe$_2$ or interface from SOC-like monolayer-MoS$_2$/NbSe$_2$ could also induce the 2D Ising superconductivity transition \cite{NbSe2, 2DNbSe2, MoS2/NbSe2}, the NbSe$_2$ layer in our present study is up to 22 UCs, which should play as a bulk crystal as well studied in previous work \cite{NbSe2}. For comparing, we also studied the temperature dependence of $H_{c2}^{ab}$ of the 22-UCs NbSe$_2$ as given in Fig. \ref{fig3}(f). Although the value of $H_{c2}^{ab}$ is comparable to that of CrOCl, the low temperatures upturn phenomenon is absent, but obeys a 3D superconductivity as the Wethamer-Helfand-Hohenberg (WHH) model, where $H_{c2}^{ab}(0)= -0.697T_c(\frac{dH_{c2}^{ab}(T)}{dT})_{T\sim T_c}$.


\section{DIFFERENTIAL RESISTANCE SPECTRA}

\begin{figure*}[b]
\centering
\includegraphics[width=0.9\linewidth]{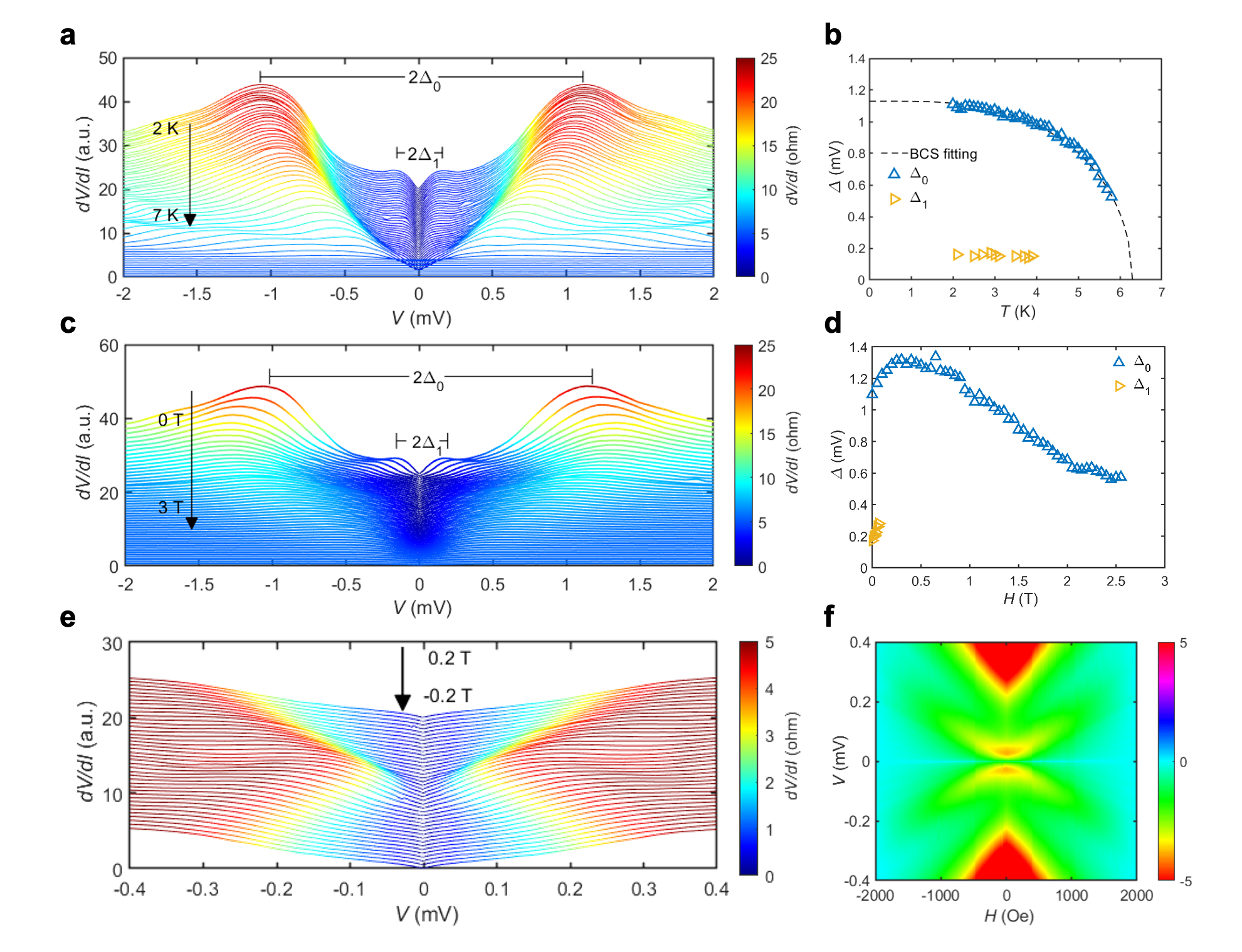}
\caption{{\label{fig4} \textbf{Differential resistance (d$V$/d$I$) spectra with respect to temperature and magnetic field.} \textbf{(a)} The $dV/dI$ spectra with temperature. Several main peaks emerge below the superconducting transition temperature of NbSe$_2$. \textbf{(b)} Dependence of superconducting gaps $\Delta_0$ and $\Delta_1$ with temperatures, extracted as the peak positions from the $dV/dI$ curves. Here, $\Delta_0$ can be fitted by the BCS model as shown in the black dash line, and the $\Delta_0$ is about 1.16 meV which is consistent with the intrinsic gap of NbSe$_2$. \textbf{(c)} The $dV/dI$ spectra with magnetic field from 0 to 3 T at 2 K. Several main peaks emerge similarly. \textbf{(d)} Dependence of superconducting gaps of $\Delta_0$ and $\Delta_1$ with magnetic fields, extracted as the peak positions from the d$V$/d$I$ curves. Note that the $\Delta_1$ is undetectable with field up to 0.2 T. \textbf{(e)} The $dV/dI$ spectra under low magnetic field from -0.2 to 0.2 T at 2 K. \textbf{(f)} The contour mapped $dV/dI$ after subtracting the 2000 Oe magnetic field background.}}
\label{fig:fig4}
\end{figure*}

We further studied differential resistance (d$V$/d$I$) for the monolayer-CrOCl/NbSe$_2$ at various temperatures and magnetic fields as shown in Fig. \ref{fig4}. For the d$V$/d$I$ spectrum at 2 K, a pair of major peaks are observed at $V_0$ = 1.16~mV which can basically considered as the superconducting gap of NbSe$_2$ ($\Delta_0$), and the gap gradually disappears above 6.3~K. Interestingly, the differential resistance has not yet reach to zero, but drops to completely zero at a considerably low $V_1$ = 0.12~mV, indicating a subgap ($\Delta_1$) in the heterostructure. Fig.\ref{fig4}(b) shows the temperature dependence of both gaps as estimated from the peaks in Fig. \ref{fig4}(a). Obviously, the temperature dependence of $\Delta_0$ obeys a typical $s$-wave superconductivity for intrinsic NbSe$_2$ based on the BCS model. For the temperature dependence of subgap, however, it is less than those of NbSe$_2$ for one order of magnitude, and vanishes at about 5.5~K which is less than the $T_c$ of NbSe$_2$ as well. The subgap should correspond to the proximity-induced superconducting gap as discussed above, which has also been reported in previous studies \cite{WTe2,BiSe}.

As applying magnetic fields, both $\Delta_0$ and $\Delta_1$ are enhanced by fields below 0.2 T as shown in Fig. \ref{fig4}(c) and \ref{fig4}(d). Surprisingly, when the field is above 0.2 T, the subgap is completely suppressed, but the $\Delta_0$ is gradually restricted by the magnetic field. The anomalous enhancement of superconductivity is probably owing to the coupling between CrOCl and NbSe$_2$. Thus, we further investigate the subgap under low-magnetic fields from -0.2 to 0.2 T as shown in the d$V$/d$I$-$V$ spectra in Fig.\ref{fig4}(e). In the presence of low magnetic fields, the subgap exhibits a notable dependence on such fields, which may be contributed by field polarization of local magnetic moments. Under a small external magnetic field, the exchange-scattering time $\tau _B \sim \mu/2\pi J^2$ ($\mu$ is chemical potential) is reduced because the exchange parameter $J$ is enhanced which weakens the impurity scattering and increases the subgap \cite{MBT}. Furthermore, the application of a magnetic field leads to a broadening of the subgap rather than a direct suppression.

Based on these results, we can basically conclude that the existence of proximity-effect-induced superconductivity within monolayer CrOCl. A direct argument on the induced superconductivity is the signal from tunneling junction as described in Fig. S7. However, once the supercurrent from tunneling junction dominates the measurement, one can observe that the superconductivity should be related to the intrinsic NbSe$_2$ completely, because the supercurrent will flow cross the junction and NbSe$_2$ layer, rather than the CrOCl layer. In this case, we can only observe a large superconducting gap related to the intrinsic NbSe$_2$, and the considerably small subgap should vanish (see Fig. S6). On the other hand, once a current electrode is connected to NbSe$_2$ directly, the supercurrent will just flow through NbSe$_2$ as well. As a result, the critical current and $dV/dI$-$V$ spectra are consistent with the intrinsic properties of NbSe$_2$ (see Fig. S6). Actually, we have studied more than 100 samples for the CrOCl/NbSe$_2$ heterostructure, among which the subgap does not always exist owing to this shortage or interface problem caused by the burrs at the electrode as introduced in the Supplementary Information. After all, the CrOCl crystals are in atomic layers, all these layers should be extremely sensitive to the substrate.

\section{The origin of proximitized superconductivity}

To understand the superconducting proximity effect in the CrOCl/NbSe$_2$ heterostructure, we carried out first-principles calculations with the slab model. Compared to the vacuum level, the Fermi level of NbSe$_2$ layers is located at the $-$5.5\,eV , which is lying in the gap of CrOCl monolayer. Thus, only the strain or defect may tune the electronic structure of CrOCl to achieve the overlap between CrOCl's energy bands and NbSe$_2$'s conducting bands. Our calculations reveal that it is very difficult to achieve proximity-effect-induced superconductivity by in-plane strain within $\pm 5\%$ magnitude. So we focus on the defect's influence about the electronic structure of the CrOCl. We note that there are two possible types of defect in the CrOCl, including Cl vacancy and Cr vacancy, and the first-principles results are present in the Fig.5 and Fig.S12. It indicates that the Cl vacancy will induce two localized energy bands as shown in the Fig.S12. These in-gap defect states have been observed in the previous experiments \cite{CrOCl_APL}, which is indeed close to the conducting bands of NbSe$_2$. But the possible proximitized superconductivity based on the localized Cl vacancy is difficult to be detected in the experiments. As for the Cr vacancy, this type of defect has been reported in the similar CrSBr system \cite{CrSBr}. Our calculated electronic bands as shown in the Fig.5 reveal that the highest valence bands of CrOCl are tuned to the Fermi level of NbSe$_2$ due to two Cr vacancies. The case of Cr-vacancy line-defect has the similar electronic phenomenon. In summary, the Cr vacancy defect can make the CrOCl's electron close to the Fermi surface of NbSe$_2$ to realize the proximitized superconductivity.

\begin{figure*}[b]
\centering
\includegraphics[width=0.8\linewidth]{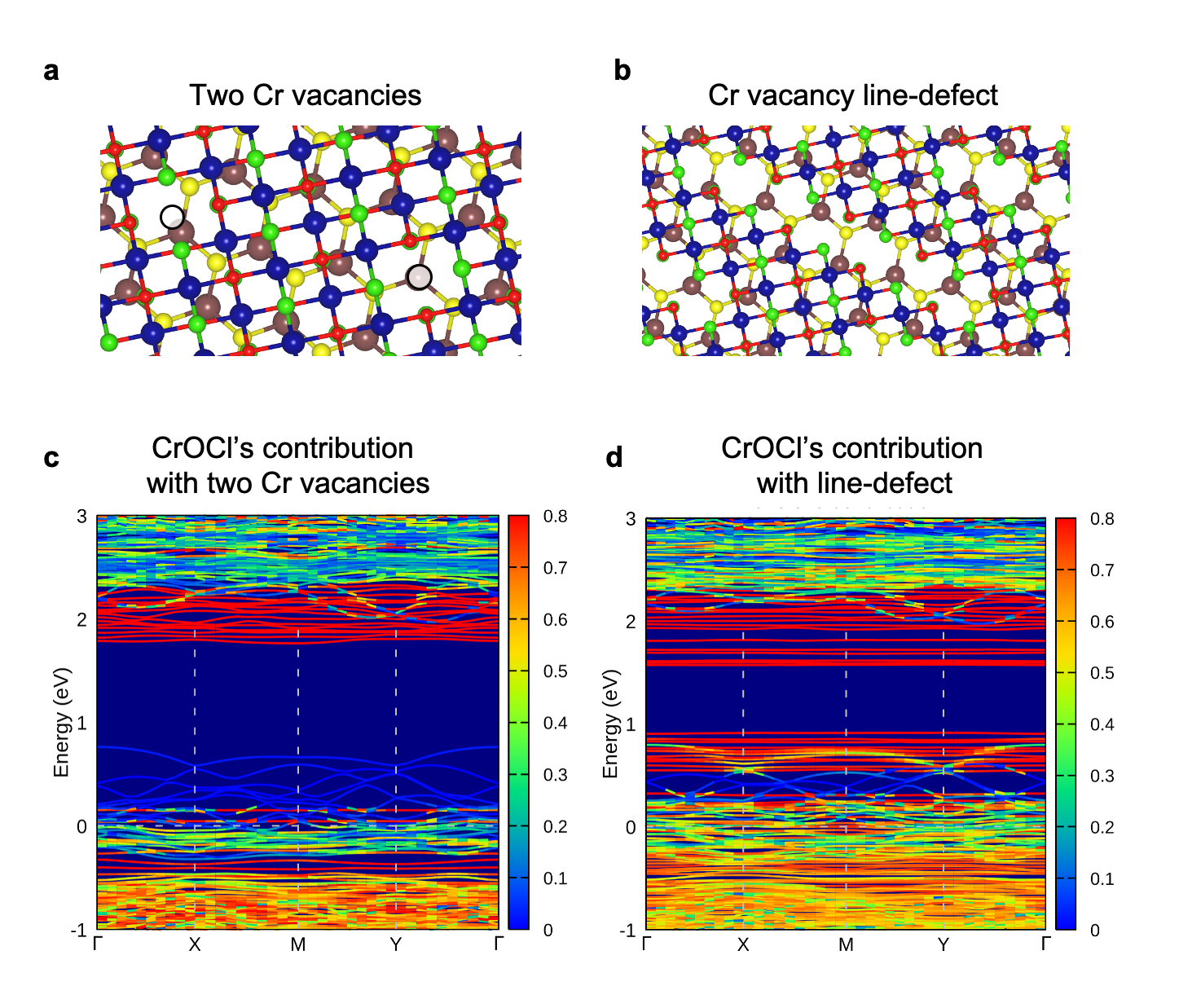}
\caption{{\label{fig5} \textbf{The origin of proximitized superconductivity in CrOCl.}\textbf{(a)} The top view of CrOCl/NbSe$_2$ heterostructure with two Cr vacancies remarked by circles. The brown, yellow, blue, red, and green balls represent Nb, Se, Cr, O, and Cl atoms, respectively. \textbf{(b)} The top view of CrOCl/NbSe$_2$ heterostructure with Cr vacancy line-defect. \textbf{(c)} The energy bands of CrOCl/NbSe$_2$ heterostructure with two Cr vacancies in the CrOCl layer. The removed Cr atoms carry opposite spins to keep the antiferromagnetism of CrOCl layer. \textbf{(d)} The energy bands of CrOCl/NbSe$_2$ heterostructure with Cr vacancy line-defect in the CrOCl layer. }}
\label{fig:figs5}
\end{figure*}

\section{CONCLUSION}
In this paper, by constructing the breaking of time reversal symmetry in magnetic insulator and $s$-wave superconductor CrOCl/NbSe$_2$ device, we have observed the superconductivity through the proximity effect. Multiple superconducting gaps have been found below the transition temperature of NbSe$_2$. Moreover, the unconventional dependence of temperature and in-plane magnetic field in induced superconductivity is also studied. The in-plane magnetic field breaks the Kramers degeneracy and FFLO state was observed which came from the spin splitting. First-principles calculations systematically illustrated the origin of proximitized superconductivity. Our results reveal the proximity-effect-induced  superconductivity with FFLO state in 2D van der Waals heterostructures and shed lights on the fascinating interaction between superconductivity and magnetism.

\section{METHODS}

\textbf{Crystal Synthesis}. The CrOCl crystals were synthesized by a solid growth technique as introduced in previous paper\cite{CrOCl}. A mixture of powdered CrCl$_3$ and Cr$_2$O$_3$ with a molar ratio of 1:1 and a total mass of 1.5 g were sealed in an evacuated quartz ampule. The ampule was then placed in a two-zone furnace, where the source and sink temperatures for the growth were set to 940 $^{\circ}$C and 800 $^{\circ}$C, respectively, and kept for two weeks. Subsequently, the furnace was slowly cooled to room temperature, and high-quality CrOCl crystals were obtained. The single crystals were also ground and studied by a powder XRD method, and the results is well consistent with previous work \cite{CrOCl}.

\textbf{Device Fabrication}. The vdW heterostructure devices were fabricated using the dry transfer technique in a vacuum condition. Thin pieces of $h$-BN, NbSe$_2$ and few-layers CrOCl were exfoliated from high-quality single crystals of $h$-BN, NbSe$_2$ and CrOCl, respectively. The NbSe$_2$ and CrOCl flakes were then picked up by polydimethylsiloxane (PDMS)/polyvinyl alcohol (PVA) polymer stacks at 85 $^{\circ}$C, transferred and aligned on the substrates in order. The $h$-BN flake was covered on the top of the heterostructure to prevent any degradation. The circuit pattern was written onto the SiO$_2$/Si substrates by Laser Direct-Write lithography system (Microwriter ML3) within standard lithography process. Next, the electrode patterns were etched by 25 nm using reaction ion etching system (RIE),  5 nm of Ti and 20 nm of Au were then grown by electron beam (E-beam) deposition technique.

Since CrOCl is the thin flake of few atomic layers, it is extremely sensitive to substrate surface quality. Especially at side of the electrodes in which some burrs often exist during the lift-off technique as discussed in Supplementary Information Fig. S5, resulting in a short pass through the NbSe$_2$ directly. To avoid this serious problem, we polished and totally cleaned the as-prepared Ti/Au electrodes. The atomic force microscope and transmission electron microscope analysis indicated that the electrodes are considerable flat, as seen in Supplementary Information Fig. S5 and Fig. S10, respectively.

\textbf{STEM Characterization}. The double Cs-corrected scanning transmission electron microscopy (STEM, Grand JEM-ARM300F) was applied to analysis the atomic structure, the microscopy was equipped with a cold field-emission gun and operated at the accelerating voltage of 300 kV. We cut the device by a focused ion beam (FIB, Grand JIB-4700F) to explore the interfaces from the cross-section view. Here the thickness of the thin specimens is around 50-100 nm. The energy-dispersive X-Ray spectroscopy (EDS) mapping was applied for the element distribution study.

\textbf{Density Functional Theory Calculations}.
We carried out first principles calculations within the framework of the generalized gradient approximation functional \cite{PhysRevLett.77.3865} of the density functional theory through employing the Vienna ab initio simulation package (VASP) \cite{PhysRevB.47.558} with projector augmented wave method \cite{PhysRevB.50.17953}. The DFT-D2 method of Grimme \cite{Grimme} is used to describe the interlayer interaction between CrOCl and NbSe$_2$ substrate.

\bigskip
\section{ACKNOWLEDGEMENT}

This research was supported in part by the Ministry of Science and Technology (MOST) of China (No. 2022YFA1603903), the National Natural Science Foundation of China (Grants No. 12004251, 12104302, 12104303, 12304217), the Science and Technology Commission of Shanghai Municipality, the Shanghai Sailing Program (Grant No. 21YF1429200), the start-up funding from ShanghaiTech University, and Beijing National Laboratory for Condensed Matter Physics, the Interdisciplinary Program of Wuhan National High Magnetic Field Center (WHMFC202124). Growth of hexagonal boron nitride crystals was supported by the Elemental Strategy Initiative conducted by the MEXT, Japan, Grant Number JPMXP0112101001, JSPS KAKENHI Grant Number JP20H00354 and A3 Foresight by JSPS.



\bibliography{SCCrOCl}

\end{document}